\documentclass[aps, prl, superscriptaddress, twocolumn,amsmath,amssymb]{revtex4-2}

\usepackage{hyperref}
\usepackage{graphicx}
\usepackage{dcolumn}
\usepackage{xcolor}
\usepackage{braket}
\usepackage{commath}
\usepackage{adjustbox}
\usepackage{comment} 
\usepackage{mathtools}

\usepackage{svg}

\newcommand{\bs}[1]{{\boldsymbol{#1}}}
\newcommand{\bk}{\bs{k}}
\newcommand{\bq}{\bs{q}}

\newcommand{\br}{\bs{r}}

\newcommand{\itbf}[1]{ {\it {\bf  #1}}}

\begin{document}

\title{Non-equilibrium pre-thermal states in a two-dimensional  photon fluid}

\author{Murad Abuzarli}
\affiliation{Laboratoire Kastler Brossel, Sorbonne University, CNRS, ENS-PSL University, Coll\`ege de France; 4 Place Jussieu, 75005 Paris, France}

\author{Nicolas Cherroret}
\email{nicolas.cherroret@lkb.upmc.fr}
\affiliation{Laboratoire Kastler Brossel, Sorbonne University, CNRS, ENS-PSL University, Coll\`ege de France; 4 Place Jussieu, 75005 Paris, France}

\author{Tom Bienaimé}
\affiliation{Laboratoire Kastler Brossel, Sorbonne University, CNRS, ENS-PSL University, Coll\`ege de France; 4 Place Jussieu, 75005 Paris, France}

\affiliation{ISIS (UMR 7006), University of Strasbourg and CNRS, 67000 Strasbourg, France}


\author{Quentin Glorieux}
\email{quentin.glorieux@lkb.upmc.fr}
\affiliation{Laboratoire Kastler Brossel, Sorbonne University, CNRS, ENS-PSL University, Coll\`ege de France; 4 Place Jussieu, 75005 Paris, France}


\maketitle
\textbf{Thermalization is the dynamical process by which a many-body system evolves toward a thermal equilibrium state that maximizes 
its entropy.
In certain cases, however, the establishment of thermal equilibrium is significantly slowed down 
and a phenomenon of  pre-thermalization can emerge. \cite{Berges2004, Kollar11,  Mallayya2019, Mori2018}. 
It describes the initial relaxation toward a quasi-steady state after a perturbation. 
While having similar properties to their thermal counterparts, pre-thermal states  exhibit a partial memory of initial conditions \cite{Berges2004, Kollar11, Mori2018, Larre2018, Martone18, Mallayya2019, Bardon-brun2020, Scoquart2020, Cherroret2021}.
Here, we observe the dynamical formation of a pre-thermal state in a non-equilibrium, two-dimensional (2D) fluid of light after an interaction quench.
Direct measurements of the fluid's first-order correlation function reveal the spontaneous emergence of long-range algebraic correlations spreading within a light-cone, providing a clear signature of a quasi steady-state strongly similar to a 2D thermal superfluid \cite{Petrov00, Greiner2001, Mora03, Hadzibabic06, Murthy2015, Clade09}. 
Detailed experimental characterization of the algebraic order is presented
and a partial memory of the initial conditions is demonstrated, in agreement with recent theoretical predictions 
\cite{Bardon-brun2020}.
Furthermore, by a controlled increase of the fluid fluctuations, we unveil a cross-over from algebraic to short-range (exponential) correlations, analogous to the celebrated Kosterlitz-Thouless transition observed at thermal equilibrium.
These results
suggest the existence of non-equilibrium precursors for thermodynamic phase transitions.
}

The relaxation dynamics of isolated many-body systems has revealed a rich variety of scenarios in the past decades \cite{Polkovnikov2011, Gogolin2016}.
While a general understanding of how a quantum system returns to its  equilibrium after a perturbation is still elusive, several intriguing phenomena have been identified.
Examples include the non-equilibrium dynamics of near-integrable systems \cite{Kinoshita2006}, the relaxation toward thermalization \cite{Trotzky12} or the spontaneous emergence of universal scaling laws \cite{Prufer2018, Erne2018, Glidden2021} following a quantum quench.
In non-equilibrium many-body systems, the phenomenon of pre-thermalization plays a central role \cite{Berges2004, Kollar11,  Mallayya2019}. Pre-thermalization is characterized by a long-lived state that relaxes to full thermalization over a much longer time scale.

In a pre-thermal state, the system retains a partial memory of its initial conditions, while showing a strong resemblance to its true thermal equilibrium state \cite{Berges2004, Kollar11, Mori2018, Larre2018, Martone18, Mallayya2019,Bardon-brun2020, Scoquart2020}. 
Experimentally, most studies of this phenomenon have been conducted in ultra-cold atomic gases:
relaxation and pre-thermalization have been observed in one-dimensional (1D) Bose gases \cite{Gring2012, Langen2013, Langen2015}, where the dynamical emergence of a pre-thermal state arises due to the presence of a trapping potential which weakly breaks integrability. Recently, signatures of a pre-thermal state were also identified in a unitary Bose gas \cite{Eigen2018}.

In parallel, fluids of light in the propagating geometry \cite{Carusotto13} have emerged as a complementary platform to study 2D Bose gases, with the observation of Bogoliubov-like dispersion \cite{Vocke15,Fontaine18, Fontaine20, Piekarski2021}, signatures of photon condensation \cite{Sun12, Santic18} and spontaneous nucleation of vortices in a photonic lattice \cite{Situ2020}.
This platform relies on the formal analogy between a laser field propagating through a nonlinear medium and the temporal evolution of a 2D quantum fluid. 
This propagating geometry is well suited to study non-equilibrium physics, since the initial state can be engineered at will using wavefront shaping techniques, 
as illustrated by recent observations of dispersive shock waves
\cite{Rothenberg1989, Xu2017, Bienaime2021, Abuzarli2021}.
Moreover, upon entering the nonlinear medium, the beam experiences a sudden change of the nonlinear refractive index, which effectively reproduces the non-equilibrium dynamics of a Bose gas after an interaction quench \cite{Steinhauer2021}.


\begin{figure*}[ht!]
\centering
\includegraphics[width=1\textwidth]{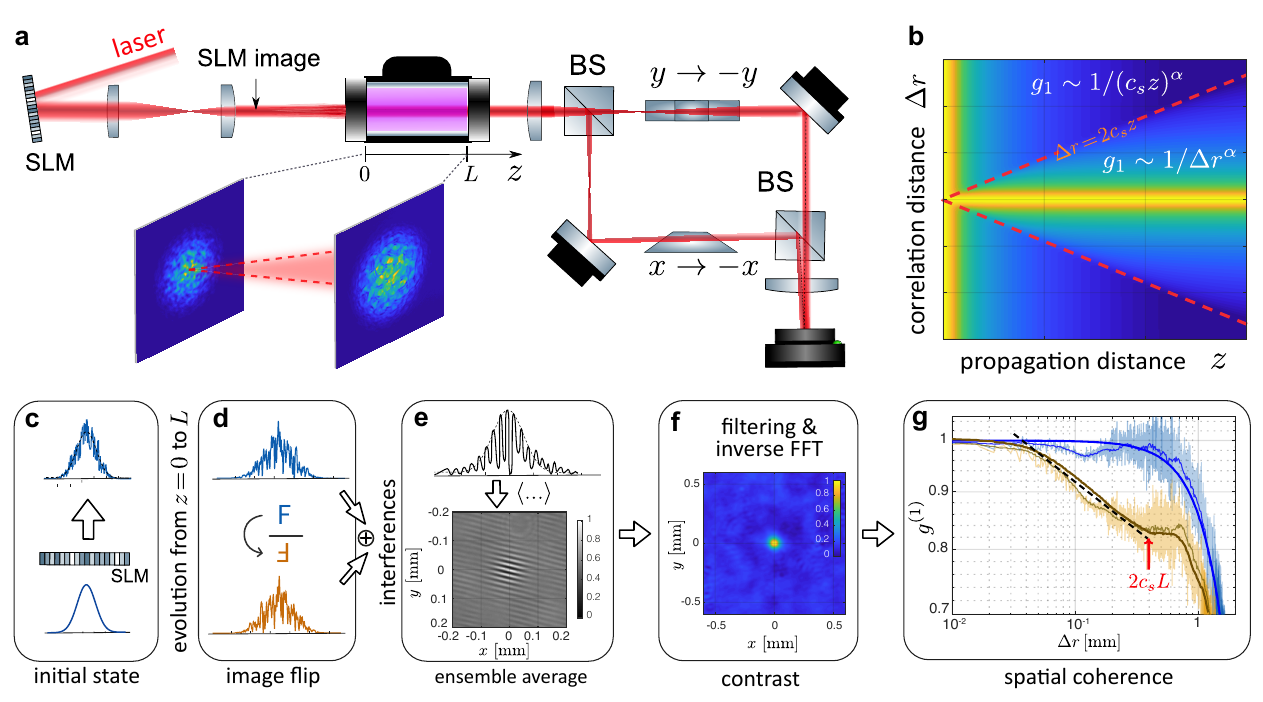}
\caption{
\label{Fig_setup}
Measurement of the coherence function demonstrating light-cone spreading of algebraic correlations in a 2D photon fluid.
(a) Sketch of the experimental setup. 
(c) An initial state is prepared using a SLM with a random phase mask superimposed on a Gaussian laser beam. This results in a weakly fluctuating field that propagates in a hot $^{87}$Rb vapor cell of length $L$.
(d) After propagation in the cell, the beam is split within a Mach-Zender interferometer and flipped using two Dove prisms.
(e) The two inverted copies interfere and the fringe visibility is recorded.
(f) The coherence function $g^{(1)}(\Delta\br=2\br)=\langle\Psi^*(\br,L)\Psi(-\br,L)\rangle$ is obtained by computing the ensemble average of the measured contrast over 2000 realizations.
(b) Scenario of pre-thermalization in 2D. Upon propagation, the initial short-range speckle fluctuations are amplified and exhibit algebraic correlations spreading within a light cone of boundary $\Delta r=2c_sL$. 
(g) Experimental $\smash{g^{(1)}}$ functions at $z=0$ and $z=L$. Light colored data are raw signals, and thin solid curves their azimuthal average. Thick solid curves are the final measurements, obtained by subtracting imperfections of the background laser.
The light cone position at $2 c_s L$ is indicated by the red arrow, and the dashed line is a guide emphasizing the algebraic decay.
}
\end{figure*}

In this Letter, we study the dynamical emergence of long-range correlations in a non-equilibrium 2D fluid of light, realized by letting propagate a laser beam through a  $10$~mm-long vapor cell of $^{87}$Rb heated to a temperature of $150 \, ^\circ$C.
Effective photon-photon interactions are achieved by tuning the laser close to resonance (detuned by $-1.5 \pm 0.1$ GHz) with respect to the $F = 2\to F'$ transition of the D2-line at $\lambda_0 = 2\pi/k_0 = 780$~nm, as described in \cite{Fontaine18,Aladjidi2022}.
Under these conditions, the vapor is self defocusing, corresponding to repulsive photon-photon interactions. 
We carefully prepare an initial state consisting of a weak random speckle field $\psi_s(\br)$ superimposed on a more intense laser beam having a wide Gaussian profile of waist $w_0=1.8$ mm and denoted by $I_t(r)=I_0\exp(-2r^2/w_0^2)$.  
The optical field impinging on the cell is of the form  $\Psi(\br,z=0)=\sqrt{I_\text{t}(r)}[1+\epsilon\psi_s(\br)]/\sqrt{1+\epsilon^2}$, where $\epsilon\ll 1$ is a dimensionless parameter quantifying the fluctuations and controlled via the SLM pattern. 
This initial state is the optical analogue of an ultra-cold  Bose gas of temperature $T_i$ (in units of the condensation temperature), which would consist of a macroscopic condensate with a condensate fraction of $1/(1+\epsilon^2)=1-T_i^2$ and a small component of thermal fluctuations with a  fraction of $\epsilon^2/(1+\epsilon^2)=T_i^2$ . 
The photon fluid effectively experiences an interaction quench at the cell entrance due to the nonlinear index change.
The cell exit plane is imaged on a camera after propagating within a balanced Mach-Zehnder interferometer.
The images of both arms are inverted with the help of two Dove prisms in a perpendicular configuration.
The coherence function $\smash{g^{(1)}}$, finally, is obtained by Fourier filtering the interference contribution,
 and averaging the field at the cell exit $\Psi(\br,z=L)$ over 2000 realizations of the speckle:
\begin{equation}
    {g^{(1)}}(\Delta\br=2\br)=\langle\Psi^*(\br,L)\Psi(-\br,L)\rangle,
\end{equation}
where $\langle\ldots\rangle$ refers to ensemble averaging [see Figs. \ref{Fig_setup}(c-f) and Supp. Mat. for details].

\begin{figure*}
\centering
\includegraphics[width=0.85\textwidth]{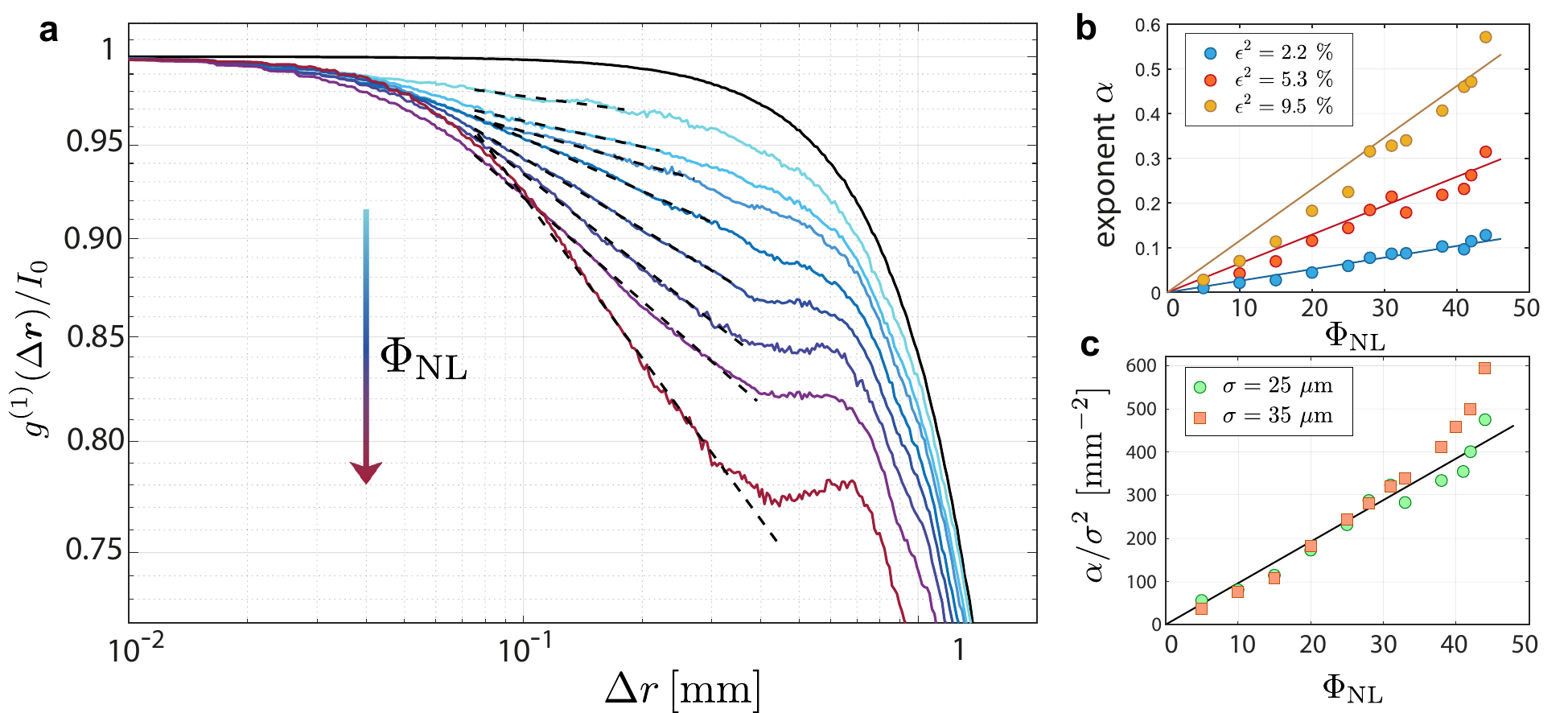}
\caption{
\label{Fig_g1_phiNL}
Direct observation of the long-range algebraic order in a 2D pre-thermal state of light.
(a) Normalized coherence function $\smash{g^{(1)}(\Delta \br)}/I_0$ vs $\Delta r$ for increasing values of the nonlinear phase $\Phi_\text{NL}$, at fixed $\sigma=25\,\mu$m and $\epsilon^2=2.2\%$. Notice the double logarithmic scale.
Dashed curves are algebraic fits to $1/\Delta r^\alpha$ in the central region.
(b) Extracted algebraic exponents $\alpha$  vs $\Phi_\text{NL}$, for three fluctuation strengths (blue: $\epsilon^2=2.2\%$, red: $\epsilon^2=5.3\%$, yellow: $\epsilon^2=9.5\%$) 
(c) Algebraic exponent $\alpha/\sigma^2$ at fixed $\epsilon^2=5.3\%$, for two different speckle correlation lengths $\sigma$ (green circles: $\sigma=25\,~\mu$m, red squares: $\sigma=35\,~\mu$m). All data points fall on the same curve, confirming the scaling $\alpha\propto \sigma^2$. 
Solid curves are linear fits to Eq. (\ref{alpha_def}).
}
\end{figure*}


Typical $\smash{g^{(1)}}$ measurements are presented in Fig. \ref{Fig_setup}(g). 
We show the raw data (light curves) obtained for a given $\Delta r$, their azymuthal average peformed by exploiting the statistical rotational invariance of $\smash{g^{(1)}}$ and, finally, the data after correction of the residual inhomogeneity of the background fluid in dark colors (see Supp. Mat.).
While, at $z=0$, the coherence function nearly coincides with the background profile $I_t(r)$ due to small initial fluctuations, at $z=L$ one observes a characteristic structure (within the background envelope) where $\smash{g^{(1)}}$ first decays algebraically  and then saturates at a constant value forming a plateau. 
A low-energy effective theory for the propagation of excitations in the vapor, assuming an homogeneous background laser, provides the dynamical scenario of 2D pre-thermalization sketched in Fig.~\ref{Fig_setup}(b) \cite{Bardon-brun2020}.
After a short propagation distance $z\sim 1/(2gI_0)$ after the quench, the initial weak speckle fluctuations are amplified, 
and a quasi long-range  order emerges spontaneously within a light cone of radius $\Delta r\sim 2 c_s z$,  where $c_s=\sqrt{g I_0/k_0}$ is the speed of sound in the vapor.
The position of the light cone can be calculated from the interaction constant $g$,  measured independently with the method of \cite{Aladjidi2022}, and is shown with a red arrow on Fig.~\ref{Fig_setup}(g). Its position agrees well with the observed onset of the $\smash{g^{(1)}}$ plateau at a propagation length $L$.

By extending the effective theory of \cite{Bardon-brun2020} to account for the Gaussian profile of the background laser, we have derived the scaling laws obeyed by the coherence function at the cell exit plane when $\epsilon\ll1$ (full derivation in the Supp. Mat.):
\begin{align}
{g^{(1)}}(\Delta\br)\propto 
I_t(r)
\begin{cases}
\left({\xi}/{\Delta r}\right)^\alpha\  \text{for} \quad \Delta r<2c_sL\\
\text{const}\ \  \ \ \ \ \text{for} \quad \Delta r>2c_sL,
\end{cases}
\label{Asymptotics_algebraics}
\end{align}
where $\xi=1/\sqrt{4 k_0 g I_0}$ is the healing length. The algebraic exponent is given by
\begin{equation}
\label{alpha_def}
\alpha
\propto
\epsilon^2\Phi_\text{NL}\sigma^2,
\end{equation}
where we have introduced the nonlinear phase $\Phi_\text{NL}=g I_0 L$ accumulated by light upon propagating through the vapor. 
Eqs. (\ref{Asymptotics_algebraics}) and (\ref{alpha_def}) emphasize a key property of the pre-thermal state: within the light cone, the fluid of light behaves like a 2D \textit{superfluid} at thermal equilibrium, characterized by a quasi long-range order \cite{Petrov00,Greiner2001, Mora03, Hadzibabic06, Clade09,Murthy2015}. 
However, as is  confirmed experimentally below, this is not a strict thermal equilibrium, since the fluid retains a partial memory of its initial state via the dependence of $\alpha$ on $\epsilon$ and $\sigma$. 

\begin{figure}
\centering
\includegraphics[width=0.95\linewidth]{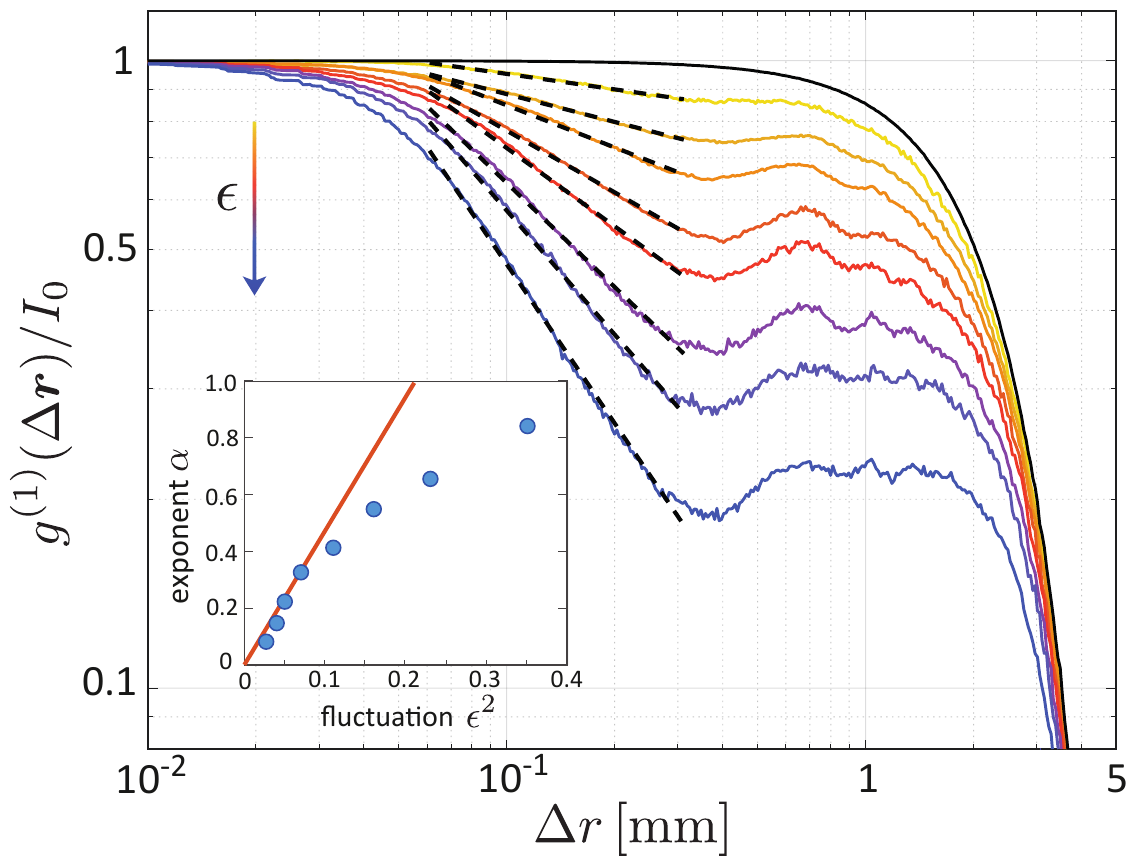}
\caption{
\label{Fig_g1_epsilon}
Impact of the initial fluctuation amplitude $\epsilon$ on the pre-thermal state.
Normalized coherence function $\smash{g^{(1)}(\Delta \br)}/I_0$ vs $\Delta r$ for increasing values of the fluctuation amplitude $\epsilon$ 
($\epsilon^2 =0, 0.027, 0.04, 0.05, 0.07, 0.11, 0.16, 0.23, 0.35$ from top to bottom).
Here $\sigma=35$~µm and $\Phi_\text{NL}=20$~rad.
Dashed curves are algebraic fits to $1/\Delta r^\alpha$ in the central region.
The inset shows the extracted algebraic exponents (blue dots), together with the prediction (\ref{alpha_def}) [with no free parameters since we set the proportionality coefficient with that of Fig. \ref{Fig_g1_phiNL}].
}
\end{figure}
\begin{figure*}
\centering
\includegraphics[width=0.85\linewidth]{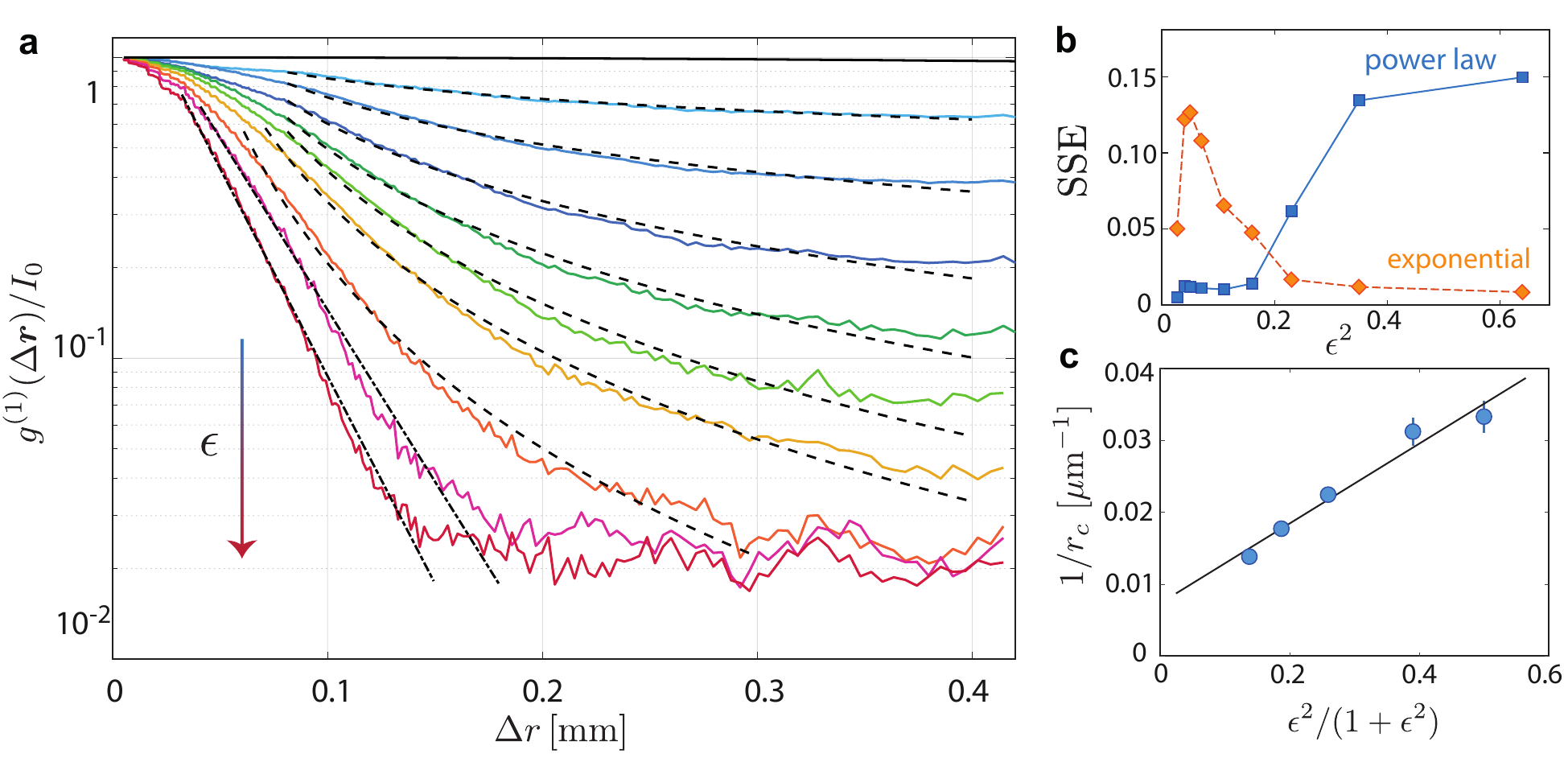}
\caption{
\label{Fig_g1_crossover}
Cross-over from algebraic to short-range (exponential) correlations at stronger  fluctuation amplitudes, with the interaction strength set to a value $\Phi_\text{NL}=44$~rad twice larger than in the previous data (notice the log scale). (a) Normalized coherence function $\smash{g^{(1)}(\Delta \br)}/I_0$ vs $\Delta r$ when increasing more significantly $\epsilon$  ($\epsilon^2 =0, 0.027, 0.04, 0.05, 0.07, 0.11, 0.16, 0.23, 0.35, 0.64$ from top to bottom). Here $\sigma=35$ $\mu$m is fixed.
(b) Sum-squared error between the experimental data and an exponential (orange) vs algebraic (blue) fit. For $\epsilon^2>0.35$ the exponential fit becomes more accurate than the algebraic fit.
(c) Rate $1/r_c$ of the exponential decay, see Eq. (\ref{eq:g1_eq}), vs the fluctuation amplitude. The linear fit (black) confirms the theroretical scaling of Eq. (\ref{eq:rc}).
}
\end{figure*}

To test the above predictions, we first show in Fig. \ref{Fig_g1_phiNL}(a) experimental coherence functions obtained for increasing values of the interaction $g\propto\Phi_\text{NL}$ at fixed $\epsilon,\ \sigma$. 
Algebraic exponents extracted from fits of $\smash{g^{(1)}(\Delta\br)}$ in the central region are shown in Fig. \ref{Fig_g1_phiNL}(b), and confirm the linear scaling of $\alpha$ with the interaction strength predicted by Eq.~(\ref{alpha_def}), at least for fluctuations $\epsilon^2$ below $10\%$.
The solid lines in Fig. \ref{Fig_g1_phiNL}(b) are linear fits to Eq.~(\ref{alpha_def}) and are used to set the proportionality coefficient of Eq.~(\ref{alpha_def}), such that there is no adjustable parameters for all other results presented in this work.
A second set of measurements, using two different correlation lengths $\sigma$, is also presented in Fig.~\ref{Fig_g1_phiNL}(c).
By normalizing the exponent $\alpha(\Phi_\text{NL})$ to $\sigma^2$, we observe that $\alpha(\Phi_\text{NL})/\sigma^2$ measured at different $\sigma$ is independent of $\sigma$, which is again in agreement with the scaling law of Eq.~(\ref{alpha_def}).

Additionally, we have investigated the behavior of $\smash{g^{(1)}}$ with $\epsilon$, setting $\sigma=35\,\mu$m and the interaction strength $\Phi_\text{NL}=20$~rad to a relatively weak value. The results are reported in Fig.~\ref{Fig_g1_epsilon}. We observe an increase of the algebraic exponent $\alpha$ with $\epsilon$.
A comparison of the extracted $\alpha(\epsilon)$ with Eq.~(\ref{alpha_def})  
is shown in the inset.
We find that the theory provides a good description of experimental results, as long as the initial fluctuations remain small, typically $\epsilon^2 \lesssim 0.25$.
In fact, deviations at higher $\epsilon$ values should not come as a surprise, since the characteristic algebraic, ``low-energy'' behavior (\ref{Asymptotics_algebraics}) 
of the coherence function is expected to hold for $\epsilon\ll 1$ only.
Indeed, in our 2D system the emergence of a pre-thermal state can be seen as a consequence of a weak breaking of translation invariance stemming from the initial speckle, a scenario for pre-thermalization put forward in \cite{Mallayya2019}.

To further characterize the non-equilibrium dynamics of our fluid of light, we have also studied the evolution of $\smash{g^{(1)}(\Delta\br)}$ up to larger values of $\epsilon$, setting a stronger interaction strength  $\Phi_\text{NL}=44$~rad. From the above discussion, one could naively expect that, upon increasing $\epsilon$, the low-energy state (\ref{Asymptotics_algebraics}) leaves room to a non-universal dynamics, where no pre-thermalization stage arises and where  $\smash{g^{(1)}(\Delta\br)}$ has no specific structure. Instead, we have experimentally observed that the coherence function smoothly turns from algebraic to exponential as $\epsilon$ is increased:
\begin{equation}
\label{eq:g1_eq}
g^{(1)}(\Delta\br)\sim\exp(-\Delta r/r_c).
\end{equation}
The cross-over from Eq.~(\ref{Asymptotics_algebraics}) to this exponential behavior is presented in the measurements of Fig.~\ref{Fig_g1_crossover}(a). 
We have confirmed it by a computation of the sum of squared estimate of errors (SSE) that  measures the discrepancy between the $g^{(1)}$ data and a fit to either Eq.~(\ref{Asymptotics_algebraics}) or (\ref{eq:g1_eq}), see Fig.~\ref{Fig_g1_crossover}(b).
Note that such an exponential decay differs from the Gaussian correlations of the initial speckle and, in that, is associated with a genuine new dynamical regime emerging  from the quench. 
Another statistical test, coefficient of determination $R^2$, is presented in the supplementary material and confirms the transition at $\epsilon^2\sim0.2$. 
We have also observed this cross-over in \textit{ab initio} numerical simulations presented in the Supp. Mat., and have found it to be a generic feature of $g^{(1)}$ in the pre-thermal regime as $\epsilon$ or $\Phi_\text{NL}$ are increased to moderate values.
This phenomenon was also previously pointed out in \cite{Bardon-brun2020}.
At a physical level, we conjecture that this algebraic-to-exponential crossover is reminiscent of the celebrated Kosterlitz-Thouless (KT) transition, which drives 2D Bose gases at thermal equilibrium from a superfluid to a normal-fluid state when the temperature is raised. 
Although out-of-equilibrium, our fluid of light  displays a very similar behavior in the pre-thermal regime. 
This unexpected phenomenon can be understood by noticing that,
at low $\epsilon$ and/or small interaction strength, the energy injected into the system during the quench is small, and so is the effective pre-thermalization ``temperature''. This results in an pre-thermal state with quasi long-range order, which can be seen as the dynamical counterpart of a 2D, equilibrium superfluid at low temperature. When $\epsilon$ and/or $g$ is increased, on the other hand, one reaches a  pre-thermal state of effectively larger temperature. The resulting fluid displays exponentially-decaying correlations, analogous to the normal phase of a 2D Bose gas above the KT temperature. 
At the cross over between the two regimes, the exponent value is $\alpha = 1.7 \pm 0.1$ (see supplementary material), in strong contrast with a KT phase transition in thermal equilibrium homogeneous systems where the exponent is 0.25 \cite{Murthy2015}. 

To gain more insight on the pre-thermal regime of exponential correlations, we have also studied the dependence of the correlation length $r_c$ of the exponential decay, see Eq. (\ref{eq:g1_eq}), on the initial fluctuation amplitude $\epsilon$. To unveil this dependence, one can take advantage of the conservation of the total energy $E_{t}=\int d\br( 1/(2k)|\nabla\psi(\br)|^2+g/2|\psi(\br)|^4)$ during the non-equilibrium evolution. Equating $E_t$ to the energy of the normal state $(\ref{eq:g1_eq})$, we obtain:
\begin{equation}
\label{eq:rc}
\frac{1}{r_c}\propto\frac{\epsilon^2}{1+\epsilon^2}\sim T_i^2.
\end{equation}
We have also confirmed this law from extensive numerical simulations presented in the Supp. Mat. Experimental values of $1/r_c$, extracted from our measurements of $\smash{g^{(1)}}$, are also shown in Fig. \ref{Fig_g1_crossover}(c). When plotted vs $\epsilon^2/(1+\epsilon^2)$, they show a good agreement with the prediction (\ref{eq:rc}).

In summary, our experimental description of a 2D fluid of light through a direct probe of its spatial coherence has revealed the dynamical emergence of algebraic pre-thermalization following an interaction quench.
Unlike previous studies involving near-integrable systems in 1D, in our case pre-thermalization emerges as a result of the weak-breaking of translation invariance after the quench. 
Our results  further point toward the existence of a cross-over from algebraic to exponential correlations in the pre-thermal regime of 2D systems, an intriguing phenomenon that we interpret as a non-equilibrium precursor of the thermodynamic KT transition. We believe that this effect 
opens exciting perspectives for further studies of non-equilibrium quantum fluids.
While a comprehensive description of 2D thermalization processes remains open, our analysis emphasizes the assets of photon fluids for its characterization, and more generally for probing the dynamics of far-from-equilibrium many-body systems.

The authors acknowledges the Agence Nationale de la Recherche (grant ANR-19-CE30-0028-01 CONFOCAL and grant ANR-21-CE47-0009 Quantum-SOPHA), 
the European Union Horizon
2020 Research and Innovation Program under Grant
Agreement No. 820392 (PhoQuS) and the Region
Île-de-France in the framework of DIM SIRTEQ for financial support.
Q. G. thanks the Institut Universitaire de France (IUF).


%
\clearpage

\section{ \Large{Methods \&\\ \vspace{0.2cm} Supplementary Material}}
%
%
%
\section{Experimental preparation and characterization of the initial state}

To produce the initial state, we start by superimposing small phase fluctuations with a Gaussian optical beam.
The phase fluctuations are generated numerically by a sequence of random numbers convoluted with a Gaussian function.
The resulting random pattern is then imprinted on the SLM (Hamamatsu, LCOS-SLM X13138, 1272$\times$1024 pixels of 12.5~$\mu$m pitch).
This leads to phase fluctuations of standard deviation  $\smash\langle\delta^2\varphi_\text{SLM}\rangle^{1/2}$ and correlation $\sigma_\text{SLM}$. A telescope  then produces a magnified image of the SLM plane (magnification factor 2.46) at a distance $\sim 10$ cm from the atomic cell.
Light propagation from the SLM plane image to the cell then converts the initial phase fluctuations into a random speckle field displaying both amplitude and phase fluctuations \cite{Goodman2008}, leading to a state of the form of Eq. (\ref{eq:initial_field}) below at the  cell entrance.

The dimensionless amplitude of speckle fluctuations, $\epsilon$, is modified via a change in $\smash\langle\delta^2\varphi_\text{SLM}\rangle^{1/2}$. 
We typically vary the latter from $2$ to $18\%$ with respect to $2\pi$. The speckle correlation length $\sigma$, on the other hand, is tuned via $\sigma_\text{SLM}$. 
In our measurements, we have used the two values $\sigma_\text{SLM}=12.5$~$\mu$m and $31.5$ $\mu$m at the mangified SLM plane, corresponding to $\sigma=25$~$\mu$m and $35$~$\mu$m at the cell entrance, respectively.
In practice, $\epsilon$ and $\sigma$ are measured 
by recording the coherence function $\smash{g_0^{(1)}}$ at the cell entrance. The profile of $\smash{g_0^{(1)}(\Delta\br)}$ is given by Eq. (\ref{eq:initial_coherence}) below, which explicitly reads:
\begin{equation}
\label{eq:fit_function}
g^{(1)}_0(\Delta\br)=
\frac{I_0}{1+\epsilon^2}\bigg[\exp\!\left(-\frac{\Delta r^2}{2w_0^2}\right)+\epsilon^2\exp\!\left(-\frac{\Delta r^2}{4\sigma^2}\right)\bigg].    
\end{equation}
We access $\epsilon$ and $\sigma$ by a direct fit of experimental data with this formula, knowing the beam waist $w_0=1.8$~mm. 
Examples of  coherence  functions measured for different values of $\smash{\langle {\delta^2\varphi_\text{SLM}}\rangle^{1/2}}$ are shown in Fig. \ref{Fig_setup}, together with their corresponding fits.
\begin{figure}
\centering
\includegraphics[width=0.9\linewidth]{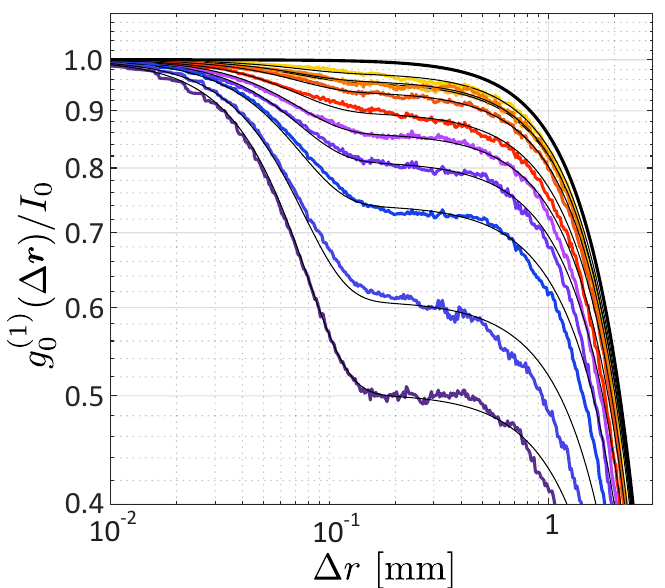}
\caption{
\label{Fig_setup}
Experimental coherence functions measured at the cell entrance, for increasing values of the SLM phase variance $\smash{\langle\delta^2\varphi_\text{SLM}\rangle^{1/2}}$. 
Solid curves are fits to Eq. (\ref{eq:fit_function}), used to access $\sigma$ and $\epsilon$. Here $\sigma=35$~$\mu$m is fixed, and $\epsilon^2 =0.027, 0.04, 0.05, 0.07, 0.11, 0.16, 0.23, 0.35, 0.64, 1$ from top to bottom. The corresponding $g^{(1)}$ functions observed at the cell exit are those shown in Fig. 3 of the main text.
}
\end{figure}

Note that, in general, knowledge of the profile $g^{(1)}_0(\Delta\br)$ alone is not sufficient to guarantee that the field fluctuations at the  cell entrance are indeed of speckle type. To verify this, we have also simulated propagations from the SLM plane to the entrance cell numerically, using the experimental  parameters. We have then analyzed the coherences $g_R=\langle\Re\psi_s(\br)\Re\psi_s(-\br)\rangle$, $g_I=\langle\Im\psi_s(\br)\Im\psi_s(-\br)\rangle$ and $g_{RI}=\langle\Re\psi_s(\br)\Im\psi_s(-\br)\rangle$ of the propagated field, and verified that $g_R=g_I$ and $g_{RI}=0$, as expected for a speckle statistics \cite{Goodman2008}.

\section{Measurement of the  coherence function}
\begin{figure}
\centering
\includegraphics[width=\columnwidth]{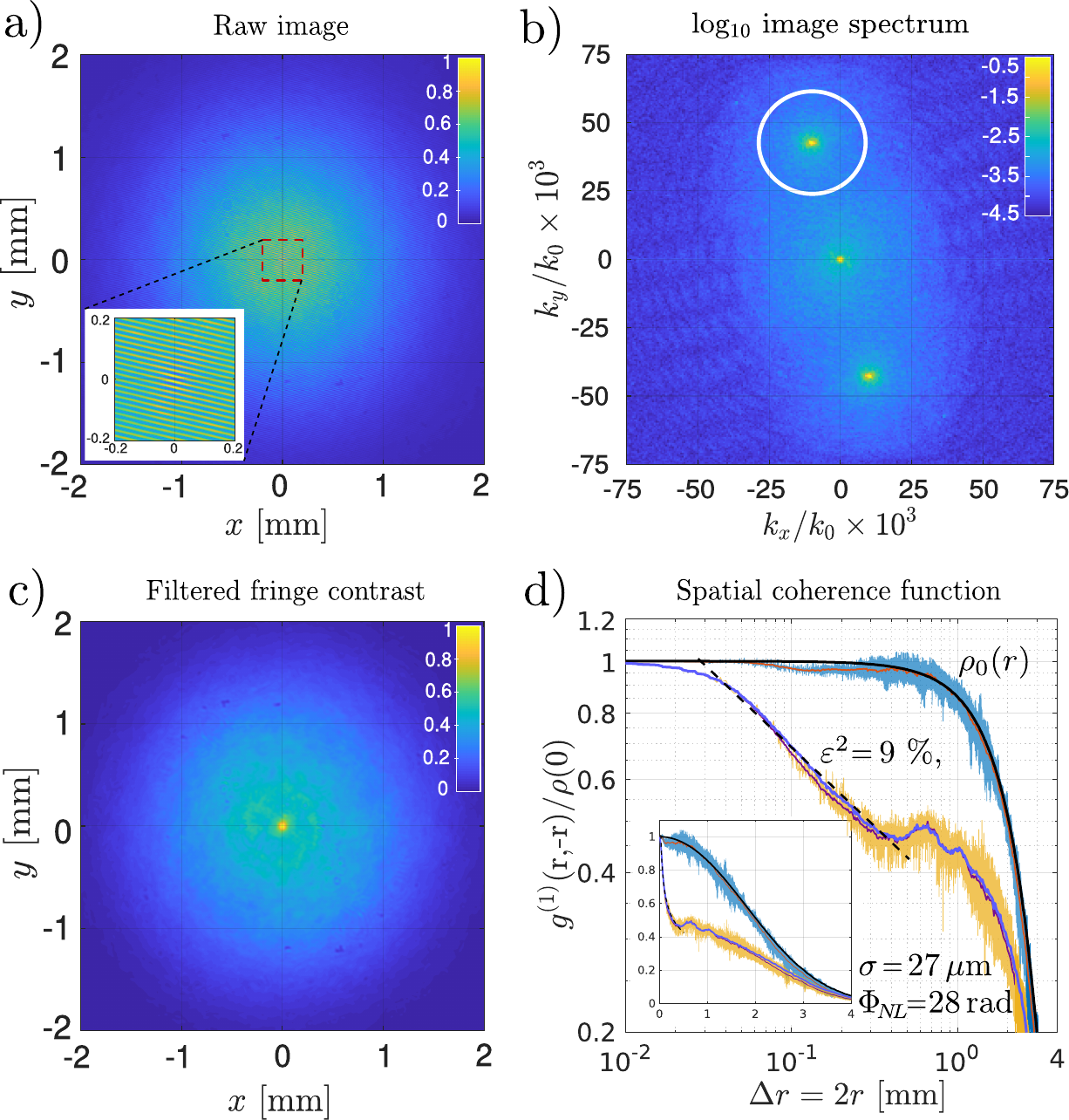}
\caption{Image processing for accessing the spatial coherence function.
a) Sum over 100 disorder realizations of the cell exit plane's interferogram image superimposed with its inverted version at an angle $\theta_r=40$~mrad.
The inset shows the zoomed square area of size 0.4~mm around the image center, on which one can distinguish interference fringes with higher contrast at the center.
b) Spatial spectrum in logarithmic scale of the image shown in a).
The white circle shows the boundary of the  filtered area.
c) Contrast map of the image a) obtained by taking the modulus of the inverse Fourier transform of the filtered spectrum of b).
d) Spatial coherence in log-log scale obtained by averaging the fringe contrast c) over the polar angle.
The top blue curves were obtained without adding phase fluctuations [$\smash{\epsilon=0,\ (\langle\delta\varphi^2_\text{SLM}\rangle^{1/2}=0}$] and hence correspond to the background density.
The bottom yellow curves were obtained for a phase standard deviation on the SLM equal to $\langle\delta\varphi^2_\text{SLM}\rangle^{1/2}=0.12\times 2\pi$ and $\sigma_\text{SLM}=2$~pixels, giving the input correlation length $\sigma=25$~$\mu$m.
The inset in d) shows the same coherence function in linear scale.}
\label{fig:ImProcess_g1}
\end{figure}

Our optical setup allows for direct measurements of the coherence function.
This measurement relies on several steps.
We first image the output plan of the non-linear medium on a camera with two imaging lenses in a 4f configuration magnifying the image by a factor 2.46. 
To directly measure the coherence function optically, we split the laser beam in two identical parts (with a 50:50 beam splitter) within a Mach-Zehnder interferometer.
Due to the radially symmetric background of our experiment, the optimal choice of configuration is to correlate pairs of points symmetric with respect to the beam center, i.e., to measure $\smash{g^{(1)}(\br,-\br)=g^{(1)}(\Delta\br=2\br)}$.
In this configuration, we have access to all relative displacements with a single frame, in which the central position corresponds to the zero relative displacement and it continuously increases as one moves away from beam center.
Experimental implementation requires flipping the images of the two arms of the Mach-Zehnder interferometer oppositely with respect to
each other.
This is done using two Dove prisms placed on rotating mounts in each interferometer arm.
Dove prism is a trapezoidal-shaped prism which flips the image with respect to a symmetry direction in the transverse plane, the latter depending on the prism's orientation angle around the optical axis.
Putting the prism's largest side horizontally at the one arm and vertically at the other arm, one ends up after recombination with a superposition of respectively vertically $(x,y)\rightarrow(x,-y)$ and horizontally $(x,y)\rightarrow(-x,y)$ flipped fields with respect to the field before the interferometer. 

Both beams are then superimposed at the camera plane with a relative angle $\itbf{\theta}_r=({\theta}_{r,x},{\theta}_{r,y})^t$,  of 40~mrad, corresponding to a relative transverse wavevector $\bk_r\simeq k \itbf{\theta}_r$.
We denote by $\br$ the spatial coordinate of any of both beams, say $\br=(-x,y)$, such that the total intensity on the camera reads:
\begin{equation}
\begin{split}
&I_{cam}(\br)\propto|\Psi(\br)+\Psi(-\br)\textrm{e}^{i\bk_r\cdot\br}|^2=\\
&|\Psi(\br)|^2+|\Psi(-\br)|^2+2\textrm{Re}\{\Psi^*(\br)\Psi(-\br) \textrm{e}^{i\bk_r\cdot\br}\}. 
\end{split}
\end{equation}
The total field $\Psi$ depends on the initial random phase $\varphi$ imprinted on the SLM.
The previous expression is ensemble averaged by acquiring multiple images, each corresponding to a different initial phase numerically calculated using a random number generating function and displayed on the SLM.
The ensemble average of the interferograms gives:
\begin{equation}
\langle I_{cam}(\br)\rangle\propto
2\rho_0(\br)\left[1 + \frac{g^{(1)}(\br,-\br)}{\rho_0(\br)} \textrm{cos}(\bk_r\cdot\br) \right],
\label{IcamResult_g1}
\end{equation}
where $\langle|\Psi(-\itbf{r})|^2\rangle=\langle|\Psi(\br)|^2\rangle=\rho_0(\br)$ and  $g^{(1)}(\br,-\br)=\langle\Psi^*(\br)\Psi(-\br)\rangle$, assuming that the coherence function has a vanishing imaginary part.

Measurement of the coherence function for a given experimental configuration ($\sigma,\epsilon,\Phi_\text{NL}$) required typically 2000 disorder realizations which were saved as 20 interferogramm images, each of them being a sum over 100 realizations.
For each experimental configuration, an additional measurement of the spatial coherence in the absence of fluctuations was performed.
This allows to identify and correct the ``static'' beam imperfections causing a small density/phase fluctuations other than the ones added by the SLM.

Given a disorder-averaged image result corresponding to Eq.~(\ref{IcamResult_g1}), and shown as an example in Fig.~\ref{fig:ImProcess_g1}a), one can retrieve the coherence function by means of Fourier filtering. 
Thanks to the relative tilt $\theta_r$ between the two interfering paths, the cosine-modulated term containing $g^{(1)}$ in Eq.~(\ref{IcamResult_g1}) appears, according to the convolution theorem, as two peaks symmetric with respect to the origin and shifted from it by a distance equal to $k_0\theta_r$. 
An example of such spectrum is shown in Fig.~\ref{fig:ImProcess_g1}b).
By filtering the area [white circle in Fig.~\ref{fig:ImProcess_g1}b)] around one of these two peaks and setting the remaining parts of the spectrum to zero, one obtains the spectrum of the complex part of the cosine term in Eq.~(\ref{IcamResult_g1}).
The absolute value of the Fourier-inverted filtered spectrum gives the fringe contrast, which is directly proportional to the $g^{(1)}$ function. 
The latter is finally normalized such that $g^{(1)}(\Delta r=0)=1$, leading to a map of the normalized coherence function as shown in Fig.~\ref{fig:ImProcess_g1}c).
This radially symmetric map is averaged at fixed $r$ over all the points sharing same radius $r$ with respect to the beam center, hence corresponding  to a two-point relative displacement equal to $2r$.
This procedure allows for a complete characterization of the coherence function.
An example of measurement is shown in Fig.~\ref{fig:ImProcess_g1}d).

\section{Calculation of the coherence function}

\subsection{Bogoliubov-de-Gennes equations}

Theoretically, the coherence function can be evaluated within a time-dependent Bogoliubov-Popov description. The case of an initial state consisting of a speckle field superposed with a uniform background was previously described in \cite{BD2020}. Here, we generalize this method to the configuration of our experiment, where the background is a spatially-dependent Gaussian field. We thus model the field impinging on the atomic cell at $z=0$ as 
\begin{equation}
\label{eq:initial_field}
\Psi_0(\br)=
\sqrt{I(r)}\left[1+\epsilon \psi_s(\br)\right]
\end{equation}
where $I(r)=I_0/(1+\epsilon^2)\exp(-2r^2/w_0^2)$ is the Gaussian background [$\br=(x,y)$, $w_0$ beam waist] and $\psi_s(\br)$ a speckle field.
The latter can be accurately modeled as a complex, Gaussian random variable with Gaussian correlation function
\begin{equation}
\langle\psi_s^*(\br)\psi_s(-\br)\rangle=\gamma(\Delta\br)\equiv   
\exp(-\Delta\br^2/4\sigma^2),
\end{equation}
where $\Delta\br=2\br$.  
The total intensity injected in the cell is, by definition,
\begin{equation}
I_\text{t}(r)=\langle|\Psi_0(\br)|^2\rangle=I_0\exp(-2r^2/w_0^2),
\end{equation}
while the coherence function of the initial state is
\begin{equation}
\label{eq:initial_coherence}
g^{(1)}_0(\Delta\br)=\langle \Psi_0^*(\br)\Psi_0(-\br)\rangle=
I(r)[1+\epsilon^2\gamma(\Delta\br)].    
\end{equation}
The pre-thermal regime described in our experiment is observed for $\epsilon\ll1$. In this limit, the initial state can be conveniently rewritten as:
\begin{equation}
\label{eq:init_cond2}
\Psi_0(\br)=
\sqrt{I_\text{t}(r)+\delta I_0(\br)}
\exp[i\theta_0(\br)]
\end{equation}
where $\delta I_0(\br)\simeq2\epsilon I(r)\psi_s^r(\br)$ and $\theta_0(\br)\simeq\epsilon\psi_s^i(\br)$, with $\psi_s^r(\br)\equiv\Re\psi_s(\br)$ and $\psi^i_s(\br)\equiv\Im\psi_s(\br)$.\newline

To compute the field $\Psi(\br,z)$ at a finite distance $z$ within the cell, we insert the Ansatz
\begin{equation}
\Psi(\br,z)=
\sqrt{I_\text{t}(r)+\delta I(\br,z)}
\exp[i\theta(\br,z)]
\end{equation}
in the paraxial wave equation
\begin{equation}
i\partial_z\Psi(\br,z)=-\frac{1}{2k}\nabla^2\Psi(\br,z)+g|\Psi(\br,z)|^2\Psi(\br,z)
\end{equation}
and linearize with respect to the fluctuations $\delta I$ and $\nabla\theta$, using the Gauge transformation $\theta(\br,z)\to \theta(\br,z)-\mu(\br) z$. The quantity $\mu(\br)$, which would correspond to a chemical potential in the context of quantum gases, here can be seen as a (local), nonlinear correction to the refractive index. It is determined from an expansion of the wave equation at zeroth order:
\begin{equation}
\label{eq:mu}
-\frac{1}{2k}\frac{\nabla^2\sqrt{I_t(r)}}{\sqrt{I_t(r)}}+g I_t(r)=\mu(\br).
\end{equation}
The intensity fluctuations and phase, on the other hand, obey the first-order Bogoliubov-de-Gennes equations
\begin{equation}
\partial_z\left(\frac{\delta I}{\sqrt{I_t}}\right)=\left(-\frac{1}{2k}\nabla^2+gI_t-\mu\right)2\sqrt{I_t}\theta
\end{equation}
and
\begin{equation}
-\partial_z\left(2\sqrt{I_t}\theta\right)=\left(-\frac{1}{2k}\nabla^2+3gI_t-\mu\right)\frac{\delta I}{\sqrt{I_t}}.
\end{equation}
We write the general solution of these equations as
\begin{equation}
\label{eq:deltaI}
\delta I(\br,z)=i \epsilon I_t(r)\sum_\nu f_\nu^-(\br)e^{-i\Omega_\nu z}+\text{c.c.}
\end{equation}
and
\begin{equation}
\label{eq:theta}
\theta(\br,z)=\frac{\epsilon}{2}\sum_\nu f_\nu^+(\br)e^{-i\Omega_\nu z}+\text{c.c.},
\end{equation}
where the Bogoliubov amplitudes $f_\nu^{\pm}(\br)$ and energies $\Omega_\nu$ has to be determined. The latter obey
\begin{equation}
\label{eq:BdG_1}
\left[-\frac{1}{2k}\nabla^2+gI_t(r)-\mu(\br)\right]f_\nu^+(\br)=\Omega_\nu f_\nu^-(\br)   
\end{equation}
and
\begin{equation}
\label{eq:BdG_2}
\left[-\frac{1}{2k}\nabla^2+3gI_t(r)-\mu(\br)\right]f_\nu^-(\br)=\Omega_\nu f_\nu^+(\br).
\end{equation}
The initial condition $\Psi(\br,z=0)=\Psi_0(\br)$, finally, imposes $\sum_\nu i(f_\nu^--f_\nu^{-*})=2\psi_s^r$ and $\sum_\nu (f_\nu^++f_\nu^{+*})=2\psi_s^i$.

\subsection{Coherence function in local density approximation}

To solve Eqs. (\ref{eq:mu}) and (\ref{eq:BdG_1})-(\ref{eq:BdG_2}), we first note that the background field is smooth, such that the kinetic pressure term in Eq. (\ref{eq:mu}) can be safely neglected, leading to $\mu(\br)\simeq gI_t(r)$.  Precisely, this approximation holds as long as $w_0\gg\xi$, with $\xi=\sqrt{1/(4gI_0 k_0)}$ the healing length. For the same reason, we can make use of a local density approximation, i.e., we write
\begin{equation}
f_\nu^\pm(\br)= f^\pm(\br,\bq)e^{i \bq\cdot\br}
\end{equation}
and $\Omega_\nu=\epsilon(\br,\bq)$,
where $f^\pm(\br,\bq)$ and $\epsilon(\br,\bq)$ are slowly varying functions of $\br$. At leading order in a gradient expansion, Eqs. (\ref{eq:BdG_1}) and (\ref{eq:BdG_2}) then read
\begin{equation}
\frac{q^2}{2k}f^+(\br,\bq)=\Omega(\br,\bq)f^-(\br,\bq)
\end{equation}
and
\begin{equation}
\left[\frac{q^2}{2k}+2g I_t(r)\right]f^-(\br,\bq)=\Omega(\br,\bq)f^+(\br,\bq).
\end{equation}
These relations are readily solved, providing
\begin{equation}
\Omega(\br,\bq)=\sqrt{q^2/2k_0[q^2/2k_0+2gI_t(r)]},
\end{equation}
which is a local Bogoliubov dispersion relation, as well as
\begin{equation}
i f^-(\br,\bq)=\psi_s^r(\bq)+i \psi_s^i(\bq)\frac{ q^2/2k_0}{\Omega(\br,\bq)}
\end{equation}
and 
\begin{equation}
i f^+(\br,\bq)=\psi_s^r(\bq)\frac{\Omega(\br,\bq)}{q^2/2k_0}+i \psi_s^i(\bq)
\end{equation}
for the Bogoliubov amplitudes.
The coherence function at finite $z$, finally, follows from 
\begin{align}
g^{(1)}(\Delta\br)=&\langle \Psi^*(\br,z)\Psi(\br,z)\rangle\\
= &I_t(r)\exp\left[-\frac{1}{2}\langle[\theta(\br,z)-\theta(-\br,z)]^2\rangle\right]\nonumber\\
&\times\exp\left[-\frac{1}{8I_t(r)}\langle[\delta I(r,z)-\delta I(-\br,z)]^2\rangle\right].\nonumber
\end{align}
Inserting Eqs. (\ref{eq:deltaI}) and (\ref{eq:theta}) into this formula, and evaluating the configuration averages, we finally obtain:
\begin{align}
g^{(1)}(\Delta\br)=&I_\text{t}(r)
\exp\bigg\{-\epsilon^2\int\!\frac{d^2\bq}{(2\pi)^2}
(1-\cos\bq\cdot\Delta\br)\gamma(\bq)\nonumber\\
&\times\bigg[1+\frac{(2gI_t(r))^2}{2\Omega^2(\br,\bq)}\sin^2\Omega(\br,\bq)z\bigg]\bigg\},
\label{g1_BG_final}
\end{align} 
where $\gamma(\bq)=4\pi\sigma^2\exp(-q^2\sigma^2)$ is the speckle power spectrum. Note that Eq. (\ref{g1_BG_final}), in particular,  reduces  to Eq. (\ref{eq:initial_coherence}) at $z=0$. At long distances $z\gg 1/(2gI_0)$, the integral over $\bq$ is dominated by low $\bq$ values.  
At intermediate scales $\xi\ll\Delta r\ll 2c_s z$, where $c_s=\sqrt{g I_0/k_0}$ is the speed of sound,
this leads to the asymptic law
\begin{equation}
\label{eq:g1_algebraic}
g^{(1)}(\Delta\br)\propto I_\text{t}(r)\left(\frac{\xi}{\Delta r}\right)^{\alpha}.
\end{equation}
The algebraic exponent is given by
\begin{equation}
\label{eq:exponent}
\alpha=2 k_0 g I_t(r)\epsilon^2\sigma^2\simeq 2k_0 g I_0\epsilon^2\sigma^2,
\end{equation}\newline
which is Eq. (3) of the main text. When $\Delta r\gg 2c_s z$, on the other hand, the integral in Eq. (\ref{g1_BG_final}) provides $g^{(1)}(\Delta\br)\sim I_\text{t}(r)(\xi/c_s z )^\alpha$ for the pre-thermalization plateau. 
\newline

\subsection{Comparison with numerical simulations}

\begin{figure}
\centering
\includegraphics[width=1\linewidth]{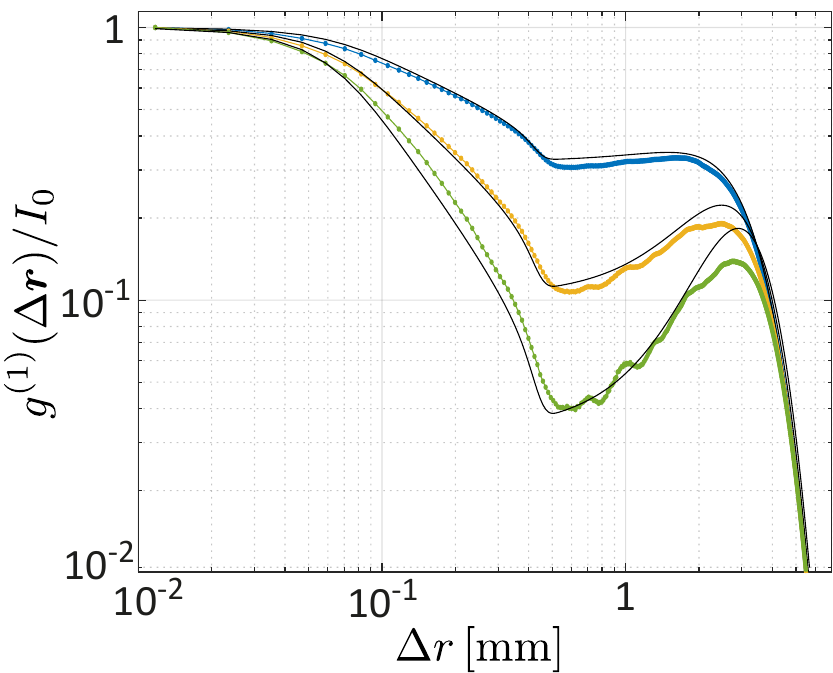}
\caption{
Coherence function at different $\epsilon$ ($\epsilon^2=0.01, 0.02, 0.03$ from top to bottom) computed from numerical simulations of the Gross-Pitaevskii equation (colored dots). Parameters of the numerical simulations are indicated in the main text. Solid curves are the theoretical prediction, Eq. (\ref{g1_BG_final}), supplemented by a fit correction factor $\beta\simeq-15$ describing phonon collisions \cite{BD2020}.
\label{g1_numerics_vs_LDA}
}
\end{figure}

To confirm the validity of eq. (\ref{g1_BG_final}), and the subsequent predictions (\ref{eq:g1_algebraic})  and (\ref{eq:exponent}) used in the main text to interpret experimental measurements, we have performed direct numerical simulations of  $\smash{g^{(1)}}$ by propagating the initial state (\ref{eq:initial_field}) with the Gross-Pitaevskii equation. This propagation relies on a standard split-step method, using a system of size $L_x=L_y=6$  discretized on a $1024\times1024$ spatial grid. The data in Fig. \ref{g1_numerics_vs_LDA} show the coherence function computed at three different $\epsilon$, at a fixed interaction $gI_0=4$, and a fixed propagation distance $z=10$ using a numerical time step $\delta z=0.04$. For the initial state, we choose a Gaussian beam waist $w_0=1.8$ and a disorder correlation $\sigma=0.025$. The results, finally, are averaged over $2500$ realizations of the initial speckle.
The numerical results in Fig. \ref{g1_numerics_vs_LDA} are shown together with the theoretical calculation, Eq. (\ref{g1_BG_final}), in which we additionally include the contribution of phonon collisions at short times, by using a phenomenological fit correction term $\beta$ within the square brackets, as explained in \cite{BD2020}. This leads to a very good agreement with the numerical data, confirming the validity of our approach and, in particular, the use of a local density approximation.

\section{Regime of exponential coherence}

\begin{figure}
\centering
\includegraphics[width=0.9\linewidth]{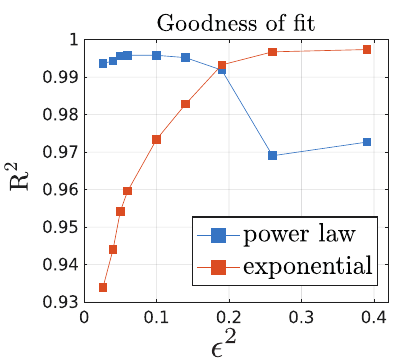}
\label{r2fit}
\caption{Goodness of the exponential or algebraic fits (coefficient of determination) as function of $\epsilon^2$. The cross over is obtained for $\epsilon^2\sim 0.2$}.
\end{figure}

As observed in the experiment, an increase of the strength of fluctuations and/or of the interaction leads to a qualitative change in the behavior of the coherence of the non-equilibrium fluid, the function $\smash{g^{(1)}(\Delta\br)}\sim\exp(-\Delta r/r_c)$ acquiring an exponential decay. 
In the main text, we provide the residual of the fits to assess which model is best suited for describing the decoherence decay. 
Another test can be used, the coefficient of determination or $R^2$.
The results are plotted in Fig. \ref{r2fit}

\begin{figure}
\centering
\includegraphics[width=0.86\linewidth]{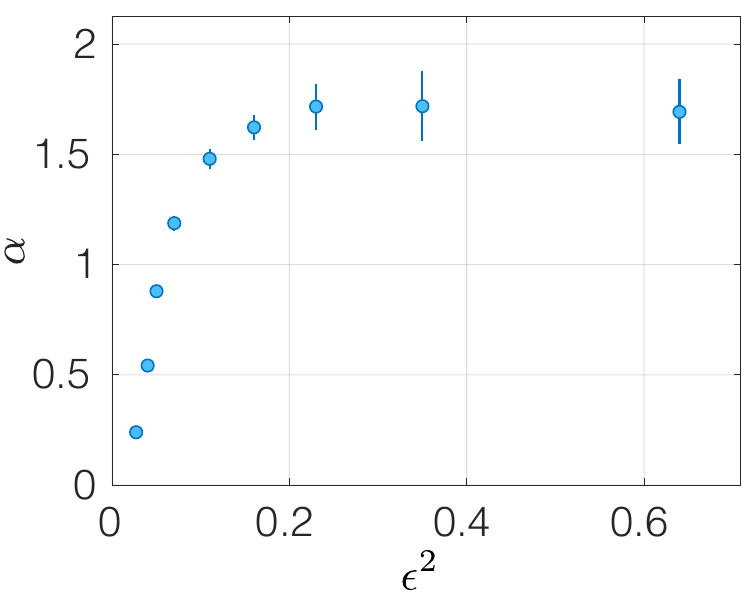}
\label{alphaBKT}
\caption{Exponent $\alpha$ as function of $\epsilon^2$. At the cross-over from algebraic decay to exponential, the value of $\alpha$ is 1.7$\pm0.1$ in strong contrast with homogeneous thermal equilibrium case where $\alpha=0.25$. Here, $\sigma=35\mu$m and $\Phi_{NL}=44$ rad.} 
\end{figure}

In Fig. \ref{alphaBKT}, we present the value of $\alpha$ as function of $\epsilon^2$. We can observe a increase of $alpha$ with $\epsilon$ at low $\epsilon$ and a saturation at the value of 1.7$\pm0.1$ for $\epsilon^2\geq0.2$

\begin{figure}[t]
\centering
\includegraphics[width=0.9\linewidth]{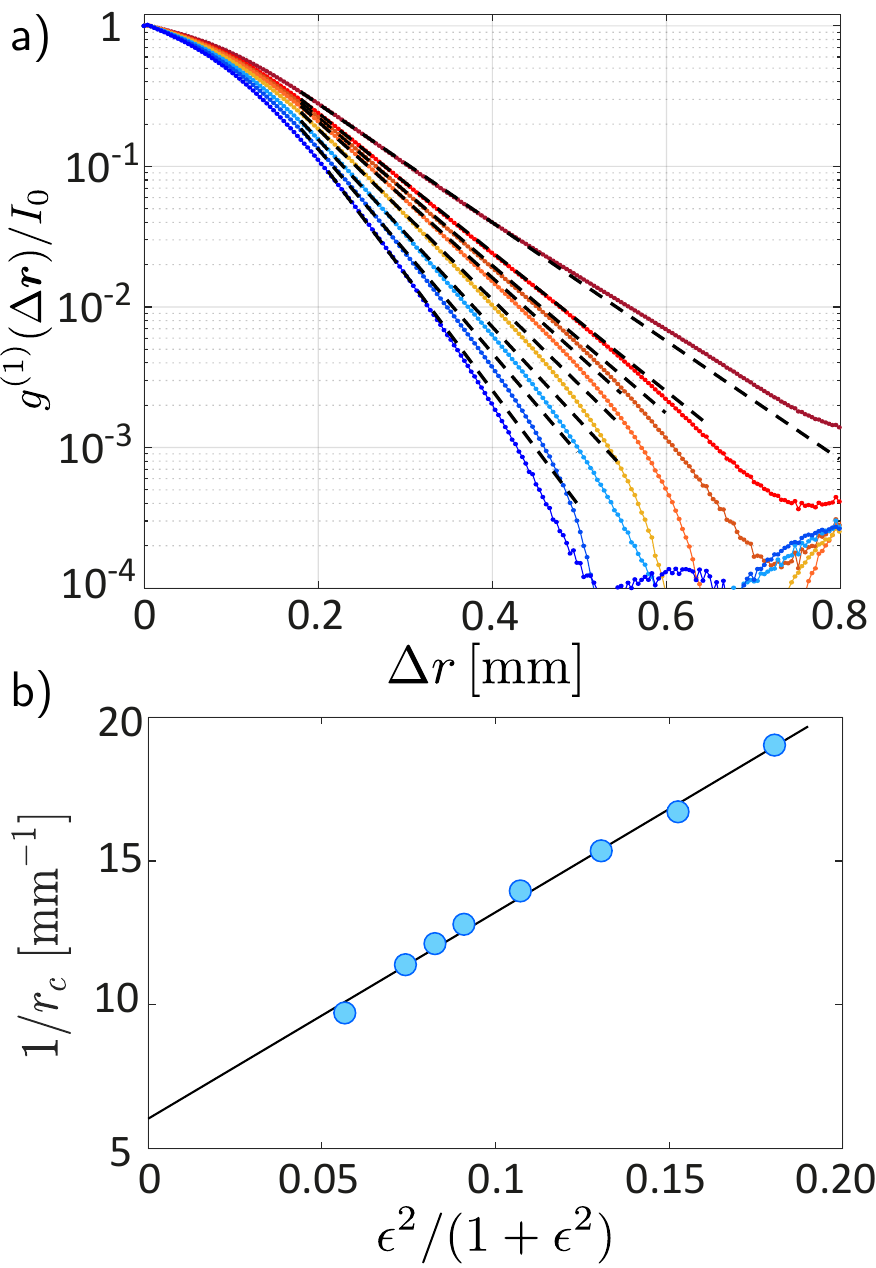}
\caption{
\label{Fig_expdecay}
a) Coherence function computed from numerical simulations of the Gross-Pitaevskii equation, using a system of size $L_x=L_y=4$  discretized on a $1024\times1024$ spatial grid. The curves correspond to
increasing values of $\epsilon$ ($\epsilon^2=0.06, 0.08, 0.09, 0.1, 0.12, 0.15, 0.18, 0.22$ from top to bottom). Here $gI_0=4.0$, and the initial state, with speckle correlation length $\sigma=35$,  has been propagated over a distance $z=44$ with a numerical time step $\delta z=0.02$. These values correspond to a nonlinear phase $\Phi_\text{NL}=gI_0 z=175$. Results are averaged over 1000 realizations of the speckle field. Dashed curves are fits to an exponential decay of the form $\exp(-\Delta r/r_c)$. 
b) Decay rate $1/r_c$ extracted from the fits. When plotted as a function of $\epsilon^2/(1+\epsilon^2)$, the points fall on a straight line.
}
\end{figure}

We have also clearly observed  this phenomenon in numerical simulations of the Gross-Pitaevskii equation, see Fig. \ref{Fig_expdecay}a).

Here the numerical simulations are performed using a uniform background (waist $w_0\to\infty$), in order to exclude any possible origin of the exponential decay coming from the  background shape, with the same interaction strength $gI_0=4.0$ as in Fig. \ref{g1_numerics_vs_LDA}
An exponential fit of numerical data provides the decay rate $1/r_c$, which is found to be proportional to  $\epsilon^2/(1+\epsilon^2)$, i.e., to the squared temperature $T_i^2$ of the initial state, see Fig. \ref{Fig_expdecay}b). This result is in full agreement with the experimental findings of the main text.

\end{document}